\documentclass[conference]{IEEEtran}

\usepackage{cite}
\usepackage{amsmath,amssymb,amsfonts}
\usepackage{algorithmic}
\usepackage{textcomp}
\usepackage{xcolor}
\def\BibTeX{{\rm B\kern-.05em{\sc i\kern-.025em b}\kern-.08em
    T\kern-.1667em\lower.7ex\hbox{E}\kern-.125emX}}
\usepackage[pdftex]{graphicx} 
\usepackage{booktabs}
\usepackage{nidanfloat}
\usepackage{hyperref}
\usepackage{enumitem}

\begin{document}

\DeclareRobustCommand*{\IEEEauthorrefmark}[1]{%
  \raisebox{0pt}[0pt][0pt]{\textsuperscript{\footnotesize #1}}%
}

\title{\vspace{0.2in}Characterizing Datasets for Social Visual Question Answering, and the New TinySocial Dataset
}
 \author{
     \IEEEauthorblockN{
         Zhanwen Chen\IEEEauthorrefmark{1},
         Shiyao Li\IEEEauthorrefmark{1},
         Roxanne Rashedi\IEEEauthorrefmark{1},
         Xiaoman Zi\IEEEauthorrefmark{1},
         Xiaoman Zi\IEEEauthorrefmark{2},
         Bryan Hollis\IEEEauthorrefmark{2}, \\
         Angela Maliakal\IEEEauthorrefmark{1},
         Xinyu Shen\IEEEauthorrefmark{1},
         Simeng Zhao\IEEEauthorrefmark{1}, and
         Maithilee Kunda\IEEEauthorrefmark{1}
     }
     \IEEEauthorblockA{\IEEEauthorrefmark{1}Department of Electrical Engineering and Computer Science, \IEEEauthorrefmark{2}Department of English, Creative Writing Program,\\ Vanderbilt University, Nashville, Tennessee, USA.  Corresponding: \{zhanwen.chen, mkunda\}@vanderbilt.edu}
 }

\maketitle

 
 
\begin{abstract}
    Modern social intelligence includes the ability to watch videos and answer questions about social and theory-of-mind-related content, e.g., for a scene in \textit{Harry Potter}, ``Is the father really upset about the boys flying the car?''  Social visual question answering (social VQA) is emerging as a valuable methodology for studying social reasoning in both humans (e.g., children with autism) and AI agents.  However, this problem space spans enormous variations in both videos and questions.  We discuss methods for creating and characterizing social VQA datasets, including 1) crowdsourcing versus in-house authoring, including sample comparisons of two new datasets that we created (TinySocial-Crowd and TinySocial-InHouse) and the previously existing Social-IQ dataset; 2) a new rubric for characterizing the difficulty and content of a given video; and 3) a new rubric for characterizing question types.  We close by describing how having well-characterized social VQA datasets will enhance the explainability of AI agents and can also inform assessments and educational interventions for people.
\end{abstract}

\section{Introduction}

\begin{figure*}[b]
    \centering            
    \includegraphics[width=\textwidth]{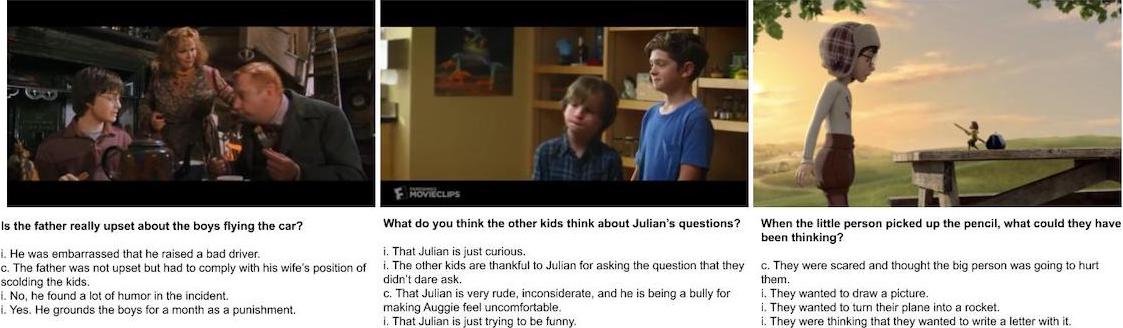}
    \caption{Sample items from our TinySocial-Crowd dataset, with questions and answers obtained through paid online crowdsourcing.  The full dataset, with video links and multiple choice questions, can be found in online supplemental material at: \url{https://figshare.com/s/81b784b5e00641f62515} \protect\cite{figshare_appendix}} 
    \label{fig:sample_question_bank}
    \vspace{-1em}
\end{figure*}

    In a scene from \textit{Harry Potter and the Chamber of Secrets}, Harry meets Ron Weasley's father at the Weasleys' breakfast table \cite{harrypotterscene}.  Hearing that the boys drove his enchanted car the previous night, the father at first excitedly asks, ``Did you really?  How'd it go?'' whereupon the mother scowls and hits him in the shoulder.  He then sternly tells the boys, ``I mean---that was very wrong indeed, boys, very wrong of you.''  The boys then share a knowing smile with each other.
    
    Is the father really upset about the boys flying the car?  Though this clip is only about 30 seconds long, most people can readily answer this question, even if they don't know anything about Harry Potter.  For example, one can make theory-of-mind (ToM) inferences about the father's internal beliefs, emotions, and desires, i.e., that he wasn't really upset and was only speaking sternly to please the mother.
    
    This task of \textit{social visual question answering} (social VQA) is just one example of how people can, seemingly effortlessly, exercise social and ToM reasoning skills.  Social VQA involves sophisticated combinations of emotion recognition, language understanding, cultural knowledge, other-agent modeling, logical and causal reasoning, and more, on top of non-social layers of comprehension about physical events.  
    Thus, social VQA is a useful paradigm for understanding human social cognition, designing practical methods for assessment and intervention, and developing AI agents with robust social reasoning skills.


    
    New social VQA datasets, i.e., collections of video clips and questions, continue to emerge across these research areas.  We suggest that considerable value can be added by characterizing such datasets along dimensions of video content and question content.  Our contributions in this paper are:
    
    \begin{itemize}[nolistsep,noitemsep]
        \item We present the TinySocial dataset that has subsets created through crowdsourcing and in-house authoring.
        \item We describe new rubrics for characterizing social content/complexity for video clips and for questions.
        \item We provide sample comparisons of three datasets: (1) TinySocial-Crowd, (2) TinySocial-InHouse, and (3) the previously existing Social-IQ dataset \cite{zadeh2019social}.
    \end{itemize}

\section{Background and Related Work}

\subsection{Social VQA for people}

    Explicit social-VQA-like interactions happen informally all the time when people discuss movies or television shows with one another.  For instance, parents often discuss videos with their children and perform what is called \textit{instructive mediation} or \textit{active mediation} by asking their children about the motives that characters might have, why certain actions are considered good or bad, etc. \cite{valkenburg1999developing}.  Instructive mediation during children's television viewing has been found to have certain positive effects on various aspects of children's behavior \cite{collier2016does}.  
    
    Implicitly, people are doing social-VQA-like reasoning every time they watch videos.  Given the amount of time that children now spend watching videos each day---e.g., in the USA, 2.5 hours daily for 8-12 year olds and nearly 3 hours daily for teenagers---such video watching is an inescapable part of social learning experiences for most children \cite{rideout2019common}.
    
    Social-VQA-like activities also appear in formal assessments \cite{heavey2000awkward,dziobek2006introducing}, stimulus materials in neuroimaging \cite{jacoby2016localizing} and behavioral \cite{black2015fiction} studies, and interventions \cite{Muller2017using}, across research on social and ToM reasoning in various populations.  For example, Movie Time Social Learning uses discussion-oriented lessons that go along with popular movies to help children on the autism spectrum learn to better identify and understand social contexts, take others' perspectives, etc. \cite{vagin2012movie}.
    
    Crowdsourcing has been used to gather materials for certain social-reasoning-oriented assessments and interventions.  One study used crowdsourcing to generate social scripts as well as common obstacles and solutions as part of a social skills instructional module for individuals with autism \cite{boujarwah2012socially}. 
    Along similar lines, another study explored if crowdsourcing could help individuals with autism find socially appropriate strategies for coping with obstacles in various social situations \cite{hong2015group}.

    \subsection{Social VQA for AI agents}
        Building on research in visual question answering (VQA), i.e., agents that can questions about static images \cite{wu2017visual}, there is a surge of interest in VQA for videos.  We use the term \textit{social VQA} to refer to video VQA where the questions target social and theory-of-mind (ToM) content, as opposed to more factual, object- or event-centric questions.
        
        Recent work in video VQA follows the pattern of training a machine learning model in a supervised fashion, using large datasets in which each training instance consists of (1) a video clip and associated features/annotations, (2) a natural language question and optionally a set of multiple choice answers, and (3) the correct answer(s).  The Social-IQ dataset uses video clips of real-world interactions from YouTube (e.g., seminars, news interviews, etc.) and presents multiple choice questions that probe social judgment, motivations and behaviors, mental states, attitudes, etc. \cite{zadeh2019social}.  Other video VQA datasets include social questions in addition to other more factual questions, such as the TVQA dataset that uses videos from six well-known TV shows \cite{lei2018tvqa}; PororoQA, that uses videos from a children's animated television show \cite{kim2017deepstory}; and MovieQA that contains questions about movies \cite{tapaswi2016movieqa}. 
        
        Outside of social VQA with videos, other AI tasks and associated datasets relevant to social and ToM reasoning include: the Theory of Mind Task dataset, that presents short textual stories and theory-of-mind-centric questions \cite{nematzadeh2018evaluating}; the Visual Beliefs dataset that presents short comic-like image sequences in which the tasks are to identify who has mistaken beliefs, and when \cite{eysenbach2016mistaken}; the Motivations dataset that contains images of people labeled with their likely motivations \cite{vondrick2016predicting}; and a dataset that contains textual, story-like descriptions of common social scripts and sociocultural norms \cite{li2012learning}.

    \subsection{Our research context}
    
    We are part of an interdisciplinary research team currently developing a new educational computer game to help middle school students on the autism spectrum improve their social and theory of mind (ToM) reasoning skills.  The game asks players to complete social VQA activities with popular television and movie clips.  An early step in this project involved putting together a small social VQA dataset to provide learning materials for the game, i.e., on the order of 100 video clips, with 6-12 multiple choice questions per clip.
    
    Our TinySocial dataset differs in several ways from datasets like Social-IQ and PororoQA.  Most notably, our dataset is intended primarily for human consumption, though we expect it may also serve as a useful test for artificial social reasoning agents.  In particular, we aimed to build a dataset that would be engaging and age-appropriate for our intended audience of middle school students, while also covering a rich and instructive variety of social and ToM-related content.
    
    As part of our participatory game design process, we invited adolescents on the autism spectrum to take part in an open-ended, social VQA activity.  Each participant watched a series of video clips with a member of our research team and then answered open-ended question about what they understood about the characters and motivations in the clip, how the clip related to their own personal experience, etc \cite{zi2020adapting}.  We also interviewed parents to share their thoughts on the everyday media experiences of these participants and areas for social growth \cite{rashedi2020opportunities}.  Several of our observations from these prior studies directly influenced the design and creation of the TinySocial dataset presented in this paper.
    
    One major observation \cite{zi2020adapting} was that clip difficulty varied based on characteristics of the clips themselves as well as of the questions asked about them.  In other words, some clips are just much harder to follow than others, but for very different reasons, from the speed of the characters' speech to the subtlety of nonverbal communication to the depth of historical themes explored.  Questions likewise vary in difficulty along many dimensions, regardless of clip difficulty; it is possible to ask very easy questions about difficult clips, and vice versa.
    
    We also debated at length about dataset creation via crowdsourcing versus in-house question authoring .  On the one hand, we felt that crowdsourcing might provide a greater diversity of question types.  On the other hand, we felt that in-house authoring might provide higher-quality questions more suitable for our target users.  So, we decided to try both methods.
    

\section{Methods for Dataset Creation}
    \subsection{Clip Selection}
        Members of our research team manually searched for video clips with rich social and emotional content on Youtube using a variety of search methods, including personal preferences, popular movies and television shows, etc.  All researchers were fully aware of the goals of our project to develop a new educational game for middle school students on the autism spectrum.  We aimed to find English-language clips that would be fun and engaging, informative about complex and relevant social and ToM content, and also age-appropriate. 
        
        We then watched each relevant video to find one- to two-minute segments containing direct or inferred social interactions rich in facial expressions, emotions, and conversations. These scenes typically involve two or more characters and exhibit a range of social/emotional content (e.g., deception, surprise, reactions, cooperation, sarcasm, discomfort, relief, frustration, shock, fear, sadness, embarrassment, joint attention, etc.).
        We also ensured that segments were relatively self-contained and understandable even if someone had no prior experience with that particular TV show or movie.
        
        
        

\subsection{In-House Question Authoring: TinySocial-InHouse}

Three members of our research team manually wrote multiple choice questions and answers for a subset of clips.  These researchers were given instructions to write a variable number of questions for each clip, as needed, and to try to cover a variety of difficulty levels and types of social and ToM content.  The researchers were also asked to include a small number of questions for each clip to gauge general comprehension, i.e., questions to assess whether the viewer was paying attention, before getting into more complex social and ToM questions.

\subsection{Crowdsourcing Question Authoring: TinySocial-Crowd}

\begin{figure}[htp]
            \centering
            \includegraphics[width=3.5cm]{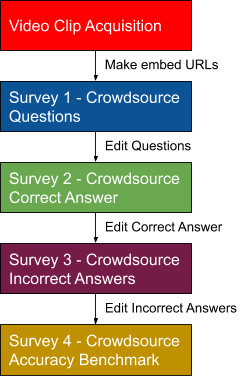}
            \caption{An overview of our crowdsourcing pipeline}
            \label{fig:survey_pipeline}
        \end{figure}

        Our Amazon Mechanical Turk (AMT) crowdsourcing pipeline consists of: (1) crowdsourcing questions, (2) crowdsourcing correct answers, (3) crowdsourcing incorrect answers, and (4) crowdsourcing accuracy benchmarks and difficulty ratings, as shown in Figure \ref{fig:survey_pipeline}. 
        Our pipeline relies heavily on the AMT Application Programming Interface (API) which allows for programmatic creation of free-form, media-rich surveys called the HTMLQuestion. We can thus create survey templates which we fill with difference video, text, and form data generated from each of our pipeline steps.  
        
        Full images of all AMT survey pages are included in this paper's online supplemental material \cite{figshare_appendix}.
        
        Because our educational game is intended for children in the United States, many of our clips are oriented in American culture and rich in cultural references unique to the United States and Canada. Therefore, we require our AMT worker registrations to be in either the United States or Canada. We also require basic worker performance measures such as having completed more than 1000 tasks and having an approval rating of at least 99 percent.

    \textbf{Step 1: Crowdsource Questions. }
        After several iterations, using a feedback-driven development process, our final crowdsourcing task design asked AMT workers to watch a short video clip write two questions in each of three categories---Comprehension, Motive, and Belief---for a total of six questions.  We provided workers first with a short description of the motivation behind our research, i.e., assessing a child's understanding of social interactions in movies, followed by 2-3 sentence descriptions of each question type, including expected difficulty for preteens.  Workers then could watch a short example clip and see example questions in each category.  Finally, we presented them with a new film clip and open text boxes in which to write new questions.  After collecting the question survey responses, we manually edit the worker-generated questions for grammar, redundancy, and quality.

    
        In earlier versions of this survey, we had tried different formulations, including question categorization (``Come up with six questions'') and difficulty categorization (``1. Easy, 2. Easy, 3. Intermediate, 4. Intermediate, 5. Difficult, 6. Difficult''). We found that requiring explicit categories increased the number of social/emotional questions compared with the other two approaches.  We also found that requiring two questions per category from each of 3 workers increased overall question diversity, as opposed to just requiring one question per category from each of 5 workers.  For example, with our initial one-question-per-category formulation using our Harry Potter clip \cite{harrypotterscene}, several workers asked why the mother hit the father on the arm instead of scolding the children.
        
        
        

    \textbf{Step 2: Crowdsource Correct Answers. }
        We use finalized questions to crowdsource correct answers. For each clip, we list each question followed by a text box for a proposed answer. We also ask workers to evaluate the quality of this question with an optional checkbox indicating a bad or unanswerable question. We assign each clip to three participants and select the best (edited) correct answer among the three and discard questions disliked by 2 or more participants. 
        
        

    \textbf{Step 3: Crowdsource Incorrect Answers. }
        We use the finalized questions and their correct answers to crowdsource incorrect answers. For a given clip, we list a single question followed by its correct answer, and three text boxes for the AMT worker to enter incorrect answers. As with correct answers crowdsourcing, we assign each survey to three participants, and then manually select the top three incorrect answers from all 9 crowdsourced options.
        
        

    \textbf{Step 4: Crowdsource Accuracy Benchmark. }
    After having the entire question bank, we proofread it before deploying a quiz for each clip consisting of its questions with one correct answer and three incorrect answers in random order. After each question, we also ask the participant to rate the difficulty of the question on a 5-point Likert scale. In addition, we also ask about their familiarity with the clip (``How many times have you watched this clip'') and demographic questions including age group, gender, ethnicity, education, native language, average daily TV/movie consumption, and years of residence in US or Canada.
    We give each survey to five workers. In this survey, we require uniqueness for workers so that we lessen the confounding factor of individual worker performance. In other words, one worker can answer at most one clip-survey whereas this was not the case with our previous surveys.
    


\section{Rubric for Characterizing Video Clips}

       To assess the level of social difficulty for each clip, we used an iterative process to develop a rubric for characterizing many different dimensions of video clips. 
       
       We first generated initial criteria, such as genre (e.g., animated; live-action), prevalence of gestures and facial expressions, perceived difficulty of language (e.g., abstract or literal), etc. To identify criteria relevant to social reasoning, our team drew on prior literature on using movie watching to enhance social reasoning for children with ASD \cite{vagin2012movie, vagin2015youcue} and previous empirical work on interventions targeted for individuals with ASD, specifically interventions focusing on improving social reasoning skills \cite{fletcher2014interventions}. 
       
       Our ratings included more complex dimensions such as, ``relevance of non-verbal cues to understanding social interactions,'' and ``perceived clarity of dialogue and perceived pace of dialogue.'' Additional dimensions which fell under the broader umbrella of social reasoning and comprehending the social content of the clips (e.g., key message of the clip; historical knowledge essential to understanding the key message in the clip) were adapted to create a clear understanding of each criterion and result in a more efficient and rapid rating system. 

        To start off, two raters independently rated four clips. 
        To ensure inter-rater reliability, our group met and discussed the two raters' individual ratings. 
        After discussing the first pass at the ratings, we decided to eliminate some criteria, including the prevalence of gestures (e.g., measured by amount of frequency on a Likert-scale of 1 to 5) and prevalence of facial expressions. These criteria were eliminated because they did not elucidate the importance of ascertaining how a specific gesture communicated something socially, or conveyed a message between characters. Through several meetings, our team realized that these criteria were superfluous, as there were other criteria that specifically assessed the intersection of body language and social content, such as ``relevance of non-verbal cues to understanding social interactions'' and ``alignment of non-verbal cues with verbal cues'' (measured by being aligned, misaligned, non-verbal only, verbal only). 
        
        \begin{figure}[t]
            \centering            
            \includegraphics[width=\linewidth]{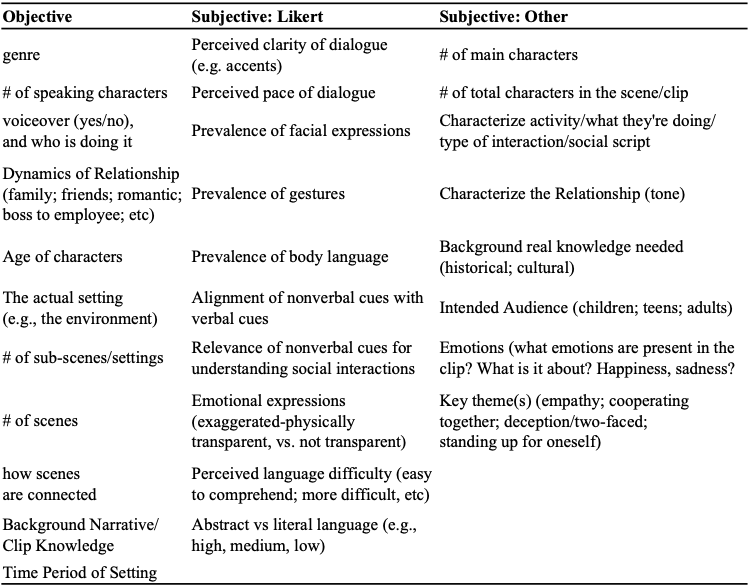}
            \caption{29 dimensions in our rubric for characterizing video clips.}
            \label{fig:rubric}
        \end{figure}
        
        During this first pass at the ratings, we also discussed needing to have a Likert-scale for some of the criterion in order to accurately and efficiently rate a criterion. For example, perceived clarity of dialogue and perceived pace of dialogue were given a 5-point scale, in part because our team categorized these criteria as more objective and thus quantifiable, rather than the more subjective criterion, such as ``describe the type of a social interaction.'' This criterion was qualitatively grounded and an open-ended response was necessary for a rater to precisely assess and describe the nature of a social interaction (e.g., initially meeting someone; parent and child experiencing tension, etc).  

        After this initial meeting, the two raters used the revised rubric to rate four new clips. The raters shared how having the Likert-scale invited them to think more precisely about how clear or unclear the dialogue was in a clip, rather than using general categories of unclear or clear. Additional discussions on having open-ended responses for some criteria, such as ``characterize the relationship'' and ``emotions,'' were also revised because the group realized that these criteria were more difficult to gauge using a 5-point Likert-scale. Identifying the emotions present in a clip using a numerical system, for instance, would not allow the raters to capture the richness and complexities of emotions (e.g., a clip showing characters feeling both sadness and confusion). 

        After this second discussion, the raters returned to the original clips and rated those again with the second iteration of the rubric, along with two new clips. The raters' results were fairly consistent and any disagreements were tabled for discussion at our third meeting, such that the raters had the space to reconcile any remaining differences. The group came to an agreement on the third and final iteration of the rubric, which contained 29 dimensions, as summarized in Figure \ref{fig:rubric}. 
        

\section{Rubric for Characterizing Questions}

A separate team of three raters developed a rubric for characterizing types of social and ToM reasoning questions.  After an initial brainstorming phase, we found that questions could be described on two separate dimensions.  The final rubric for rating questions thus consists of two separate lists of characteristics.  Within each list, the rater can select one or more items, i.e., the items are not exclusive.

Revisions of this rubric were also performed in an iterative process, though the modifications after the initial brainstorming phase were fairly minor, e.g. adding one or two rating possibilities under each dimension.

\smallskip \noindent
A) What were the \textbf{types of clues} from the clip that were relevant for answering the question?  \textbf{(1)} Words being said; \textbf{(2)} tone/volume/timbre/pitch of what is said (prosody); \textbf{(3)} facial expressions; \textbf{(4)} gestures/posture/body language; \textbf{(5)} physical actions / objects / events; \textbf{(6)} environment / scene; \textbf{(7)} artistic effects (background music, canned laughter, sound effects).

\smallskip \noindent
B) What \textbf{type of knowledge or reasoning process} did the question draw upon?
\begin{enumerate}[nolistsep,noitemsep]
    \item "factual" (i.e. was the viewer paying attention)
    \item emotions
    \item relationships: surface relationship (siblings, etc.)
    \item relationships: functional relationship / abstract roles (e.g. conflict-starter, etc.)
    \item relationships: power/social hierarchy
    \item context (scene or characters): setting
    \item context (scene or characters): personality / characteristics
    \item reasoning: motivation/intention/goal
    \item reasoning: belief
    \item reasoning: general attitude
    \item reasoning: specific attitude / communicative thing
    \item reasoning: prediction
    \item reasoning: figure of speech / sarcasm / etc. (surface meaning vs deeper meaning)
\end{enumerate}

\section{Results and Discussion}

\begin{figure*}[htp]
            \centering            
            \includegraphics[width=0.75\linewidth]{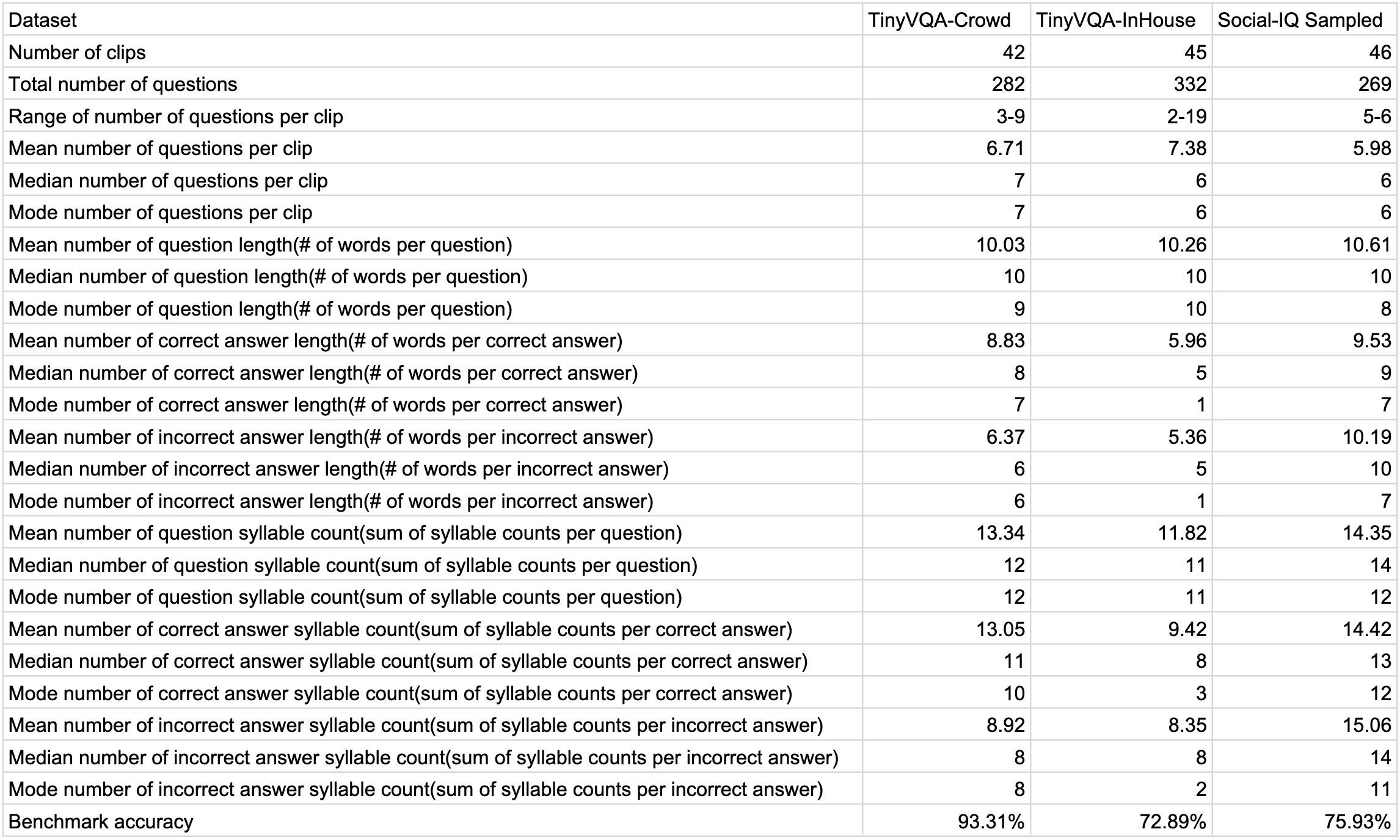}
            \caption{TinySocial and Social-IQ Dataset Statistics}
            \label{tab:dataset_stats}
        \end{figure*}

We performed comparisons of three datasets: TinySocial-Crowd, TinySocial-InHouse, and the previously existing Social-IQ dataset \cite{zadeh2019social}.  For some analyses, we randomly selected a subset of 50 clips from the Social-IQ dataset.  All datasets, including links to Youtube videos and multiple choice questions and answers, are linked in the online supplemental material \cite{figshare_appendix}.  
Due to space constraints, we present only brief summaries of our observations across the three datasets, as shown in Table \ref{tab:dataset_stats}.  A more complete presentation of all figures and results is given in the online supplemental material \cite{figshare_appendix}, including results of the clip and question ratings.

    \subsection{Potential Answering Heuristics}
        How well could intelligent agents do on our datasets using simple answering heuristics? To answer this question, we analyze a few potential strategies including random guessing, picking the shortest answer, and picking the longest answer.
        
        We created a sample from the Social-IQ dataset for a fair comparison. Due to the difference in size, we used all 42 videos from our dataset and randomly sampled 42 from the Social-IQ dataset. Additionally, because our dataset has 1 correct answer and 3 incorrect answers for each question while Social-IQ has multiple correct answers and multiple incorrect answers, we also randomly sampled one correct answer and three incorrect answers from each question in Social-IQ. We refer to this new sample as the Transformed Social-IQ dataset. 
        
\begin{figure*}[htp]
                    \centering            
                    \includegraphics[width = \textwidth]{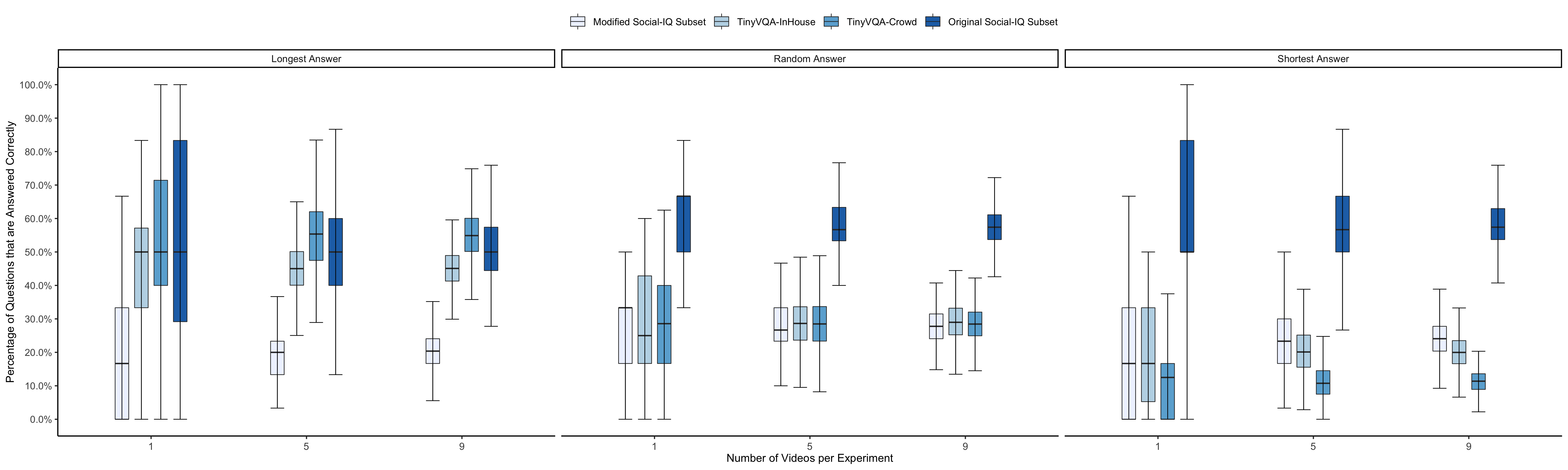}
                    \caption{Strategies Simulations Comparison. Note that Original Social-IQ has 4 correct and 3 incorrect answers per question while modified Social-IQ has 1 correct and 3 incorrect answers.}
                    \label{strategies}
        \end{figure*}

        We simulated the three answering strategies by applying each strategy on subsamples of the clips with different sample sizes ($N=1, 3, 5, 7, 9, 11$). For each sample size, we simulate applying each strategy 1000 times. The accuracy distribution over each sample size for each strategy is shown in Figure \ref{strategies}.
        
        
\section{Conclusion}
    We have created a methodology and dataset for social VQA that we expect will be useful for both cognitive and AI research. This dataset serves as a question bank for social reasoning on film clips, and we also provide extensive methods for characterizing the social content in clips and in questions. 
    
\section{Acknowledgments}
We thank the anonymous reviewers for their helpful feedback on this paper.  The research reported here was supported by the Institute of Education Sciences, U.S. Dept. of Education, through Grant R324A180171 to Vanderbilt University. The opinions expressed are those of the authors and do not represent views of the Institute or the U.S. Dept. of Education.

\bibliographystyle{IEEEtran}
\bibliography{IEEEabrv,main.bib}



\end{document}